\documentclass[superscriptaddress, preprint, longbibliography]{revtex4}

\usepackage[pdftex]{graphicx}

\usepackage[utf8]{inputenc}
\usepackage[T1]{fontenc}
\usepackage{mathptmx}
\usepackage{etoolbox}

\usepackage{amsmath}
\usepackage{amssymb}

\makeatletter
\def\@email#1#2{%
 \endgroup
 \patchcmd{\titleblock@produce}
  {\frontmatter@RRAPformat}
  {\frontmatter@RRAPformat{\produce@RRAP{*#1\href{mailto:#2}{#2}}}\frontmatter@RRAPformat}
  {}{}
}%
\makeatother
\usepackage[utf8]{inputenc}

\begin{document}

\title{A Tape-Peeling Model for Spatiotemporal Pattern Formation\\
by Deformed Adhesives}
	\author{Keisuke Taga}
	\email{tagaksk@akane.waseda.jp}
	\affiliation{Department of Physics, School of Advanced Science and Engineering, Waseda University, Tokyo 169-8555, Japan}

	\author{Yoshihiro Yamazaki}
	\affiliation{Department of Physics, School of Advanced Science and Engineering, Waseda University, Tokyo 169-8555, Japan}
\date{\today}
\begin{abstract}
We propose a new model 
for pattern formation in peeling of an adhesive tape 
based on the equation of motion for the displacement of deformed adhesives in the peel front.
The spatiotemporal patterns obtained from the model 
are consistent with those from previous models and experiments.
Moreover, dynamical and statistical properties of the patterns are investigated.
\end{abstract}
\maketitle

\begin{figure}[h]
    \centering
    \includegraphics[width=12cm]{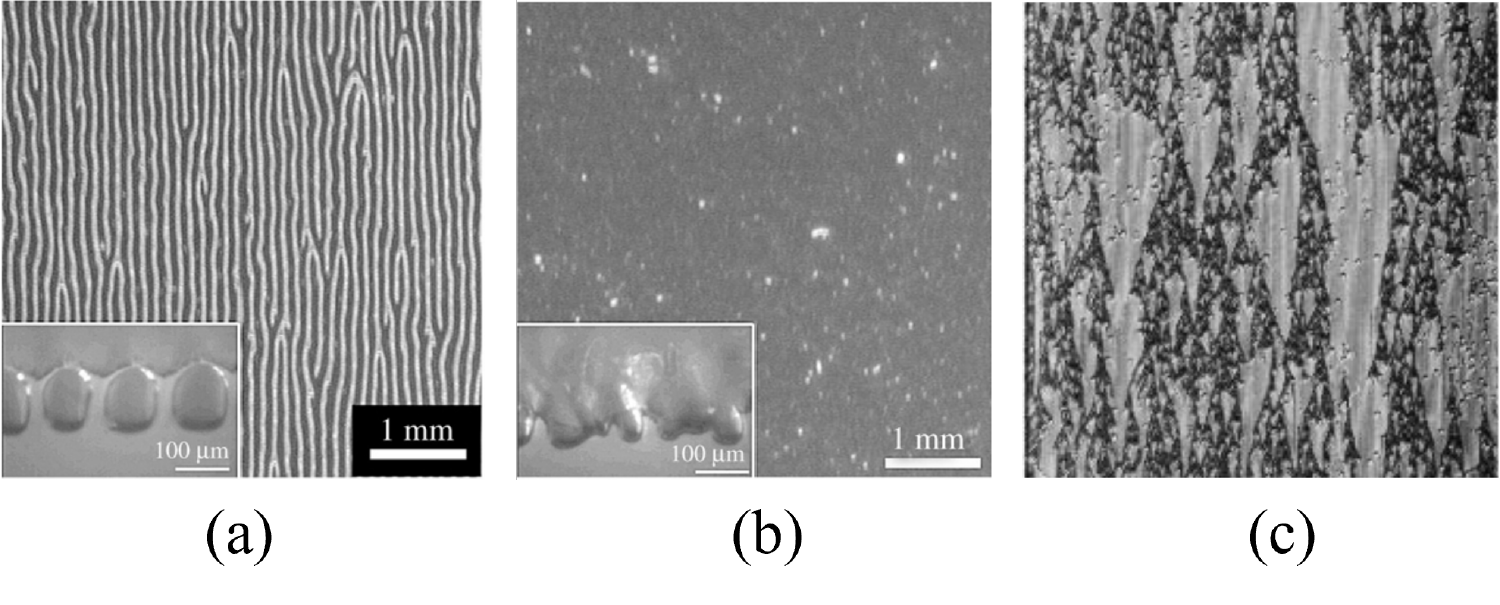}
    \caption{Peel states (a) with and (b) without 
        tunnel structures. Peeling proceeds from top to bottom. The small image at the bottom left in each figure shows the enlarged image of the peel front. (c) A fractal coexistence pattern. 
        Black and white regions represent the peel states 
        without and with the tunnel structures, respectively. 
        The actual size of the figure is $25\mathrm{mm}\times 25\mathrm{mm}$.}
    \label{peelingTrace}
\end{figure}

\textit{Introduction}: 
Peeling adhesive tapes is a common daily activity, 
and if we look closely at the tape after peeling, 
we may find interesting problems in nonlinear dynamics and statistical physics.
Previously one of the authors has investigated dynamical and 
statistical properties of spatiotemporal patterns 
composed of two different types of deformed adhesives in peeling 
\cite{Yamazaki2002-uw,Yamazaki2004-hw,yamazaki2006pattern,yamazaki2012spatiotemporal}.
The difference originates from whether a characteristic structure, 
{\it tunnel structure}, exists or not 
as shown in Figs.\ref{peelingTrace} (a) and (b).
It has been found that the peel speed and stiffness of the system are the main factors for controlling the formation of the spatiotemporal patterns and the formation of the tunnel structure.

Figure \ref{peelingTrace} (c) shows a typical pattern 
in the case of high stiffness.
Peeling proceeds from top to bottom.
Black and white regions show peel states without and with the tunnel structure, respectively.
In the previous experimental studies, it has been known that there exists a peel speed region 
where a coexistence pattern such as shown in Fig.\ref{peelingTrace}(c) is formed.
In this speed region, the following properties of the spatiotemporal patterns 
have been found \cite{Yamazaki2002-uw,Yamazaki2004-hw,yamazaki2006pattern,yamazaki2011bistable,yamazaki2012spatiotemporal,ohmori2019comments}.
(i) As the peel speed increases, 
peel state with tunnel structure becomes hard to emerge, 
and then the ratio of the white region decreases in the patterns.
(ii) Interchange of connectivity between black and white regions occurs by changing the peel speed.
(iii) Furthermore, the ratio of the white region in the coexistence pattern becomes a monotonically decreasing function of peel speed.
\cite{Yamazaki2004-hw,yamazaki2006pattern,ohmori2019comments}.

To reproduce this pattern formation, several models have been proposed so far 
\cite{Yamazaki2004-hw,yamazaki2006pattern,yamazaki2011bistable,yamazaki2012spatiotemporal,ohmori2019comments,kumar2008intermittent}.
Among these models, we now comment on the model proposed in ref.\cite{Yamazaki2004-hw,yamazaki2006pattern,yamazaki2011bistable}.
In this previous model, the state variable for the stability of the tunnel structure is introduced, 
and the peel front is considered to consist of bistable units described by the state variable.
Then, the equation of motion for the peel front 
and the time evolution of the state variable for each unit are constructed.
The asymmetric local interaction is introduced between the nearest neighbor units according to the experimental observation. 
This asymmetricity between the units causes such an effect that once the tunnel structure collapses, then the neighboring tunnel structures tend to collapse.
The previous model with the asymmetric interaction succeeds to describe the dynamics of peeling adhesive tapes. However, the relation between the state transition process and the adhesive deformation process is not clear. 
If we consider the mechanical dynamics of the adhesive deformation in peeling, the asymmetric interaction contradicts the law of action-reaction, namely, we need to construct another model for adhesive deformation in the peel front.

In this letter, we propose a new model where the asymmetricity of the two states is introduced not from a spatial interaction 
but from different local dynamics of the two peel states. 
Here we focus on the case of high stiffness and discuss dynamical and statistical properties of the model.

\begin{figure}[h]
    \centering
    \includegraphics[width=\linewidth]{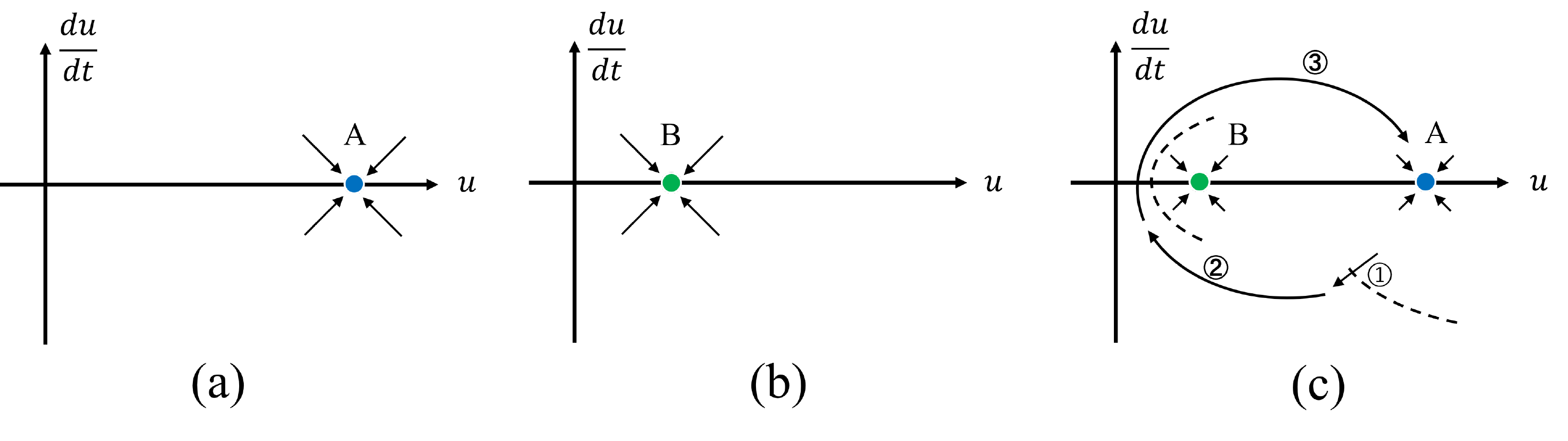}
    \caption{(Color online) Illustration of the assumed phase space dynamics.\\
    (a) Small $V$. The system has only one stable fixed point as the state A. 
    (b) Large $V$.  The system has only one stable fixed point as the state B.
    (c) Intermediate $V$. The system has two stable fixed points; the states A and B. If the tunnel structure in the state A collapses with a large perturbation as \textcircled{1}, then the state goes to the state B as \textcircled{2}. And at last, the state comes back to the state A (\textcircled{3}).}
    \label{fig:mechanism}
\end{figure}

\textit{Modelling}: 
It has been experimentally found that the peel front with the tunnel structure has larger deformation of adhesives than that without it~\cite{yamazaki2006pattern}.
As in the previous model, the peel front is divided into discrete units with the size of the tunnel structure.
The difference in adhesive deformation of each unit can be described by its displacement $u$, or $u_i(t)$ at $i$th unit in the peel front at time $t$.
Large $u$ and small $u$ correspond to peeling 
with the tunnel structure and without it, respectively.
Hereafter, the peel states with and without tunnel structure are referred to 
as the states A and B, respectively.
For the time evolution of $u$, 
we consider the Newton's equation of motion and 
assume the phase-space dynamics illustrated in Fig.~\ref{fig:mechanism} is realized in the system.
For the slow or fast peeling, the system has only one stable fixed point at large $u$ or small $u$,
 which corresponds to the state A or B as shown in Figs.~\ref{fig:mechanism}(a) and (b). 
And at the intermediate peel speed, the tunnel structure in the state A adjacent to the unit in the state B tends to collapse,
however, the tunnel structure can regenerate after a meanwhile.
This dynamics seems to resemble a threshold firing dynamics and its dynamics in the phase space can be illustrated as shown in Fig.~\ref{fig:mechanism}(c).

Based on the above consideration, the following equation is proposed for each unit:
\begin{align}
    \frac{d^2 s}{d t^2}=-\frac{d U(u)}{d u}-b\frac{d s}{d t}+f\left(u,\frac{du}{dt}\right), 
    \label{eq:general model}
\end{align}
where $s=u+Vt$ shows the length of peeled tape, and we assume $s = 0$ at $t=0$. 
$U$ is a double-well potential for the bistability of the two peel states, 
$b$ is viscosity of adhesives, and $V$ corresponds to the peel speed. 
The last term on the right-hand side shows a negative dissipation, this term is introduced from the following consideration.

After the state B occurs as the fixed point, where the system is at between Fig.~\ref{fig:mechanism}(a) and (c), there are 2 bottoms of the double-well potential corresponding to the state A and B. In this situation, the state A is dominant and the state B is unstable or unstable with a perturbation. Namely, we consider there is an additional dissipation term $f$ around the state B.
As we assume that the state B is a fixed point, thus, $f$ should be a function of $\frac{du}{dt}$ and also, it is expected that $f$ changes the positive and negative with the sign of $\frac{du}{dt}$, namely, $f\sim \frac{du}{dt}$ is the lowest order approximation.
Also this term effects mostly around the state B. Thus, $f$ is a single modal function of $u$ and it can be approximated with the second order function of $u$.
Thus, the lowest order approximation of such function is $f\sim (a-c(u-d)^2)\frac{du}{dt}$ where $a$ is the intensity, $c$ is the width of this term and $d$ is set around the $u$ corresponding to the state B. However, this function does not work well with our numerical simulation. So we also consider a physical assumption that this term converges $0$ with $|u|\to \infty$ and we introduce the dissipation term as $f=\frac{a}{1+c(u-d)^2}\frac{d u}{d t}$, where $a, c$ and $d$ have similar meanings as described above.

For the dynamics of $u$ instead of $s$, we can rewrite eq.(\ref{eq:general model}) as 
\begin{equation} 
    \frac{d^2 u}{d t^2}=-\frac{d U(u)}{d u}-b\left(V+\frac{d u}{d t}\right)+\frac{a}{1+c(u-d)^2}\frac{d u}{d t}. 
\label{eq:model2}
\end{equation}
On the right-hand side of eq.(\ref{eq:model2}), 
the terms $\displaystyle -\left(\frac{d U(u)}{d u}+bV \right)$ 
and $\displaystyle \left( -b + \frac{a}{1+c(u-d)^2} \right) \frac{d u}{d t}$ 
correspond to elastic and viscous forces, respectively. 

Furthermore, we add the following symmetric spatial interaction terms 
in eq.(\ref{eq:model2}): 

\begin{equation} 
    D_1(u_{i+1}+u_{i-1}-2u_i)+D_2\left(\frac{d u_{i+1}}{d t}+\frac{d u_{i-1}}{d t}-2\frac{d u_{i}}{d t}\right), 
    \label{eq:model3} 
\end{equation} 
where $D_1$ and $D_2$ are positive constants. 
These two terms also express the effect of the viscoelasticity of adhesives, $D_1(u_{i+1}+u_{i-1}-2u_{i})$ from the elasticity and $D_2\left(\frac{d u_{i+1}}{d t}+\frac{d u_{i-1}}{d t}-2\frac{d u_i}{d t}\right)$ from the viscosity.

\textit{Results}: 
Based on eq.(\ref{eq:model2}) with eq.(\ref{eq:model3}), 
we calculated the following equation,
\begin{equation}
    \begin{split}
    \frac{d^2u_i}{d t^2}=&-3(u_i-1)^2(u_i-2)-\left(V+\frac{d u_i}{dt}\right)+\frac{2}{1+20(u_i-1)^2}\frac{d u_i}{d t}\\
    &+(u_{i+1}+u_{i-1}-2u_i)+0.1\left(\frac{d u_{i+1}}{d t}+\frac{d u_{i-1}}{d t}-2\frac{d u_i}{d t}\right),\quad i=1,\ldots,N.
    \end{split}
    \label{eq:model4}
\end{equation}
$u_i=0$ corresponds to the state where no adhesive deformation occurs.
And we set $u_i\approx 2$ and $\approx 1$ for the states A and B, respectively.
In numerical calculation, in order to take relaxation of adhesive deformation due to spatial inhomogeneity of adhesives into account, 
we reset $u_i(t)=0$ with a probability $p=0.001$ per unit time.
As the initial condition, we set $u_i(0) = 0$ and $\frac{d u_i}{d t} = 0$. 
Periodic boundary condition was adopted.
The 4th Runge-Kutta method was used with $dt=0.01$ in time.
The discretized system size was $N=1000$.

\begin{figure}[h]
    \centering
    \includegraphics[width=1\linewidth]{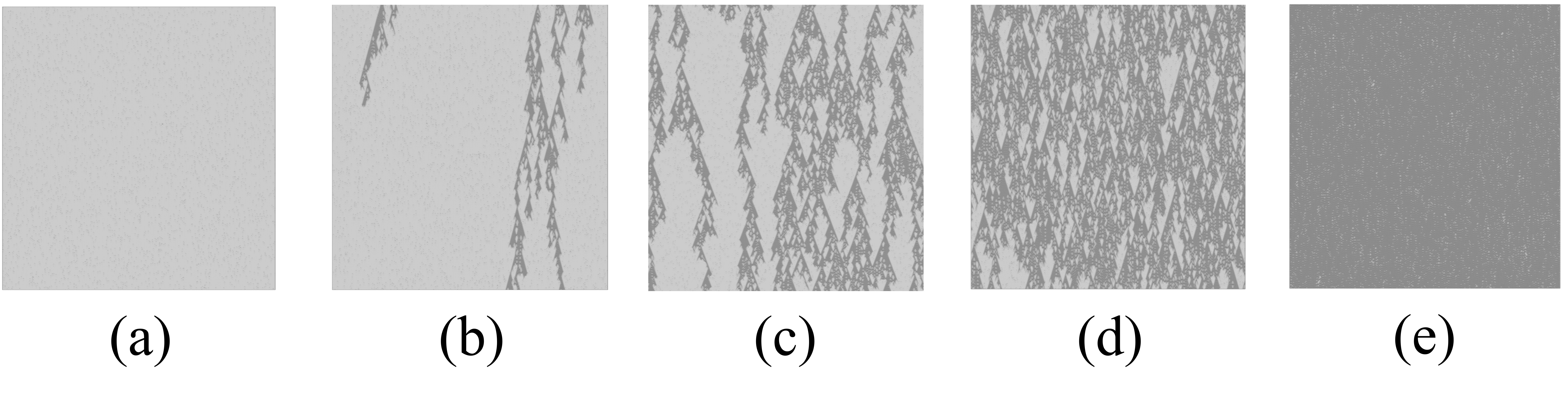}
    \caption{Typical spatiotemporal patterns obtained by eq.(\ref{eq:model4}). 
    Time proceeds from top to bottom ($500 \leq t \leq 1200$). 
    (a) $V=0.29$, (b) $V=0.30$, (c) $V=0.309$, (d) $V=0.32$, (e) $V=0.50$.}
    \label{fig:modelPattern}
\end{figure}

Figure \ref{fig:modelPattern} shows the typical spatiotemporal patterns 
obtained by the numerical calculation. 
The black and white regions correspond to the states B and A, 
respectively.
There are five cases with different values of $V$.
It is found that as $V$ increases 
the ratio of white regions (the state A) decreases.
And interchange of connectivity between black and white regions 
is confirmed.
These results are consistent with those in the previous study\cite{Yamazaki2004-hw}.

Here the following scaling properties for the coexistent patterns are focused on: 
(i) the cumulative distribution $F(\geq s)\sim s^{-\xi}$ 
where $s$ is the size of the white clusters, 
(ii) their standard deviations of the height $h(s)\sim s^{\nu_\parallel}$ 
and the width $w(s)\sim s^{\nu_\perp}$, 
and (iii) the fractal dimension $D$ of the spatiotemporal patterns 
of the white regions. 
Our numerical results for $\xi$, $D$, $\nu_\parallel$, 
and $\nu_\perp$ are shown in Table.\ref{table:exponents}. 
For comparison, this table also has their values obtained 
in the previous studies \cite{yamazaki2011bistable,ohmori2019comments}.
It is found that our present result is consistent with the previous results.

\begin{table}[h]
    \caption{The estimated values of the exponents.}
    \label{table:exponents}
    \centering
    \begin{tabular}{lcccc}
      \hline
              & $\xi$  & $D$  &  $\nu_\perp$  &  $\nu_\parallel$  \\
      \hline
      eq.(\ref{eq:model4}) at $V = 0.309$ & 0.78 & 1.82 & 0.63 & 0.45 \\
      phase-field model~\cite{yamazaki2006pattern} at $V=2.2$, $\tau=1$ & 0.91 & 1.78 & 0.61 & 0.31\\
      model A at $r=0.275$~\cite{yamazaki2011bistable} & 0.85 & 1.70 & 0.58 & 0.41 \\
      model B at $r=0.181$~\cite{ohmori2019comments} & $0.84(\pm 0.04)$ & $1.61(\pm 0.01)$ & $0.58(\pm 0.02)$ & $0.41(\pm 0.03)$ \\
      model C at $r=0.037$~\cite{ohmori2019comments} & $0.81(\pm 0.05)$ & $1.62(\pm 0.01)$ & $0.58(\pm 0.03)$ & $0.41(\pm 0.034)$ \\
      peeling at $V = 0.48\mathrm{mm}$~\cite{yamazaki2011bistable}  &  0.85 & 1.70 & 0.59 & 0.45 \\
      \hline
    \end{tabular}
\end{table}

\textit{Discussion}: Here we comment on difference 
between the previous model\cite{yamazaki2006pattern} and the present model.
The previous model has an asymmetric spatial interaction 
which expresses the situation of the state transition where the tunnel structures (in the state A) adjacent to a part of the peel front in the state B tend to collapse
\cite{yamazaki2011bistable}.
In the present model, we assume the local dynamics of the peel front as shown in Fig.~\ref{fig:mechanism} and introduce the local dissipation 
as a function of $u$ instead of the previous asymmetric spatial interaction.

\begin{figure}
    \centering
    \includegraphics[width=\linewidth]{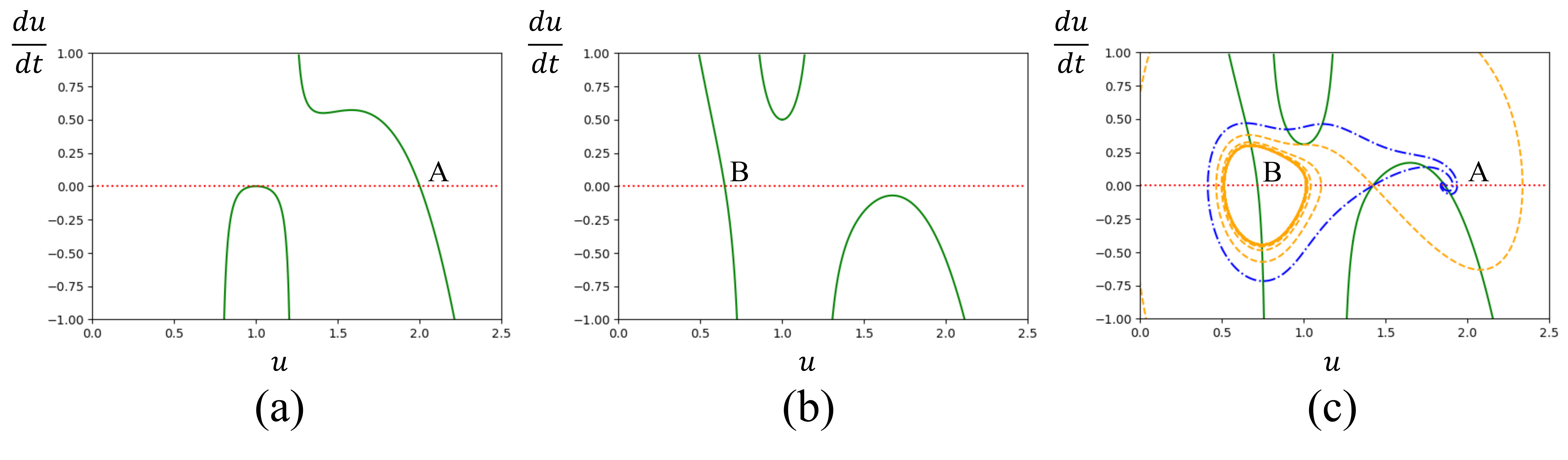}
    \caption{(Color online) Phase spaces of eq.~(\ref{eq:model4}) without spatial interaction. Red (dotted) and Green (solid) lines are the nullclines for $u$ and $du/dt$. Orange (dashed) and Blue (dash-dot) lines are stable and unstable manifolds of the saddle point.\\
    (a) $V=0$. The system has only one stable fixed point as the state A. 
    (b) $V=0.5$.  The system has only one stable fixed point as the state B.
    (c) $V=0.309$. The system has two fixed points as the state A and B, and a threshold firing dynamics is realized.}
    \label{fig:phaseSpace}
\end{figure}

Figure \ref{fig:phaseSpace} shows the nullclines and stable and unstable manifolds of eq.(\ref{eq:model4}) without the spatial interaction terms.
For the slow or fast peeling, the system has only one stable fixed point at large $u\approx 2$ or small $u\approx 1$, which corresponds to the states A or B in Figs.\ref{fig:phaseSpace}(a) and (b).
And the assumed dynamics at the intermediate speed is also realized as shown in Fig.~\ref{fig:phaseSpace}(c); $\textcircled{1}$ if the state A crosses over the stable manifold then $\textcircled{2}$ the state goes to the state B and $\textcircled{3}$ the state comes back to the state A.
We consider that such a threshold firing mechanism is essential for the pattern formation of peeling adhesive tapes, although we have to experimentally verify the validity of each term in the present phenomenological model in the future.

This bifurcation occurs in the following way. With $V$ increases from Fig.\ref{fig:phaseSpace}(a), 
there appears a saddle point and an unstable fixed point around the state B by a saddle-node bifurcation. 
Then there occurs Hopf bifurcation and an unstable limit cycle appears around the state B. 
The amplitude of the limit cycle increase with $V$ and the basin of the state B becomes larger, 
namely, the stability of the region of the state B improves as shown in Fig.\ref{fig:phaseSpace}(c), 
then the spatiotemporal pattern is obtained.
At last, homoclinic bifurcation occurs and just after that, a saddle-node bifurcation occurs and the fixed points around the state A annihilate as shown in Fig.\ref{fig:phaseSpace}(b).

\begin{figure}
    \centering
    \includegraphics[width=0.3\linewidth]{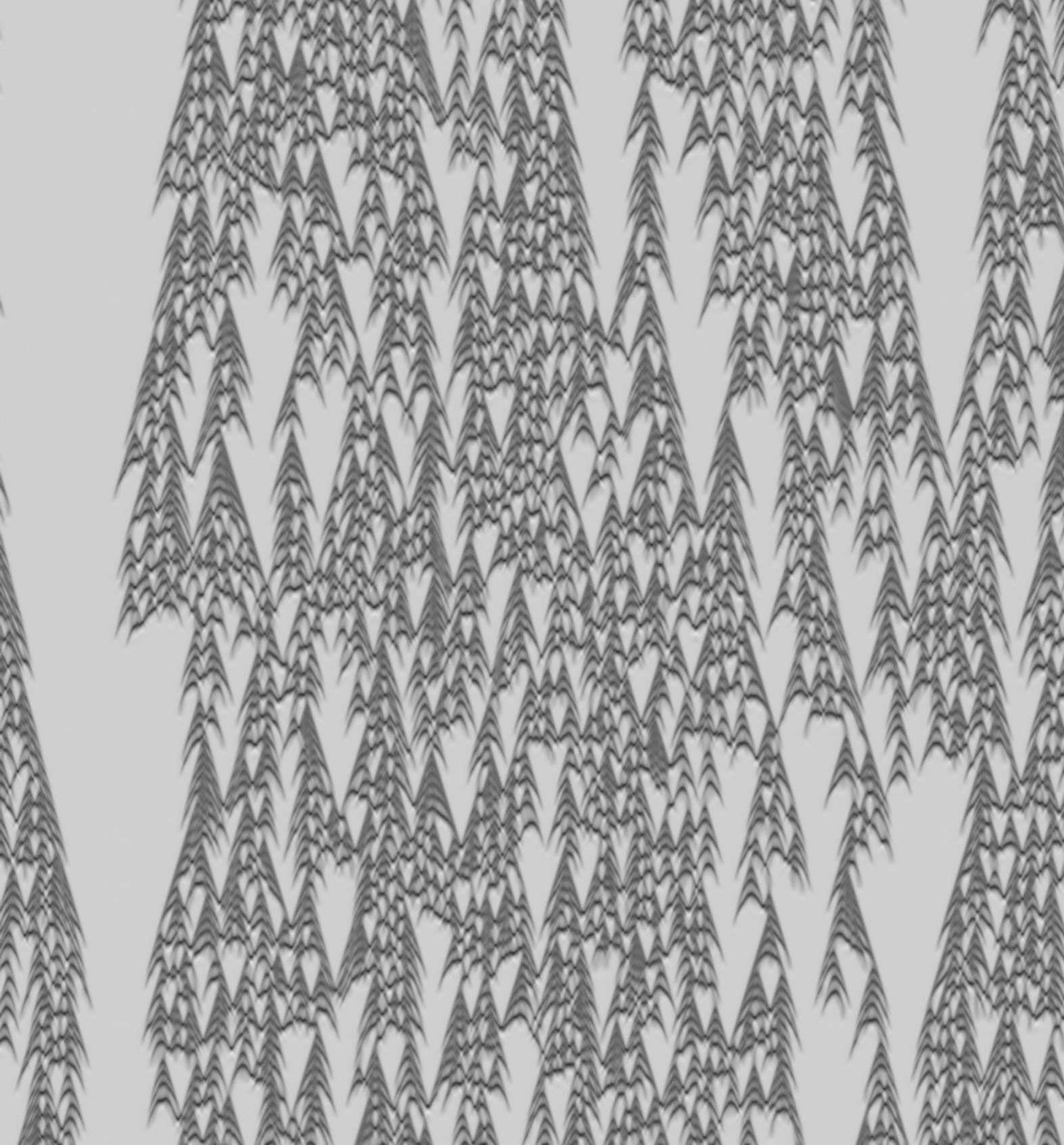}
    \caption{A spatiotemporal pattern obtained 
    from eq.(\ref{eq:model2}) with eq.(\ref{eq:model3}), 
    $\frac{d U}{d u}=0.2(u+0.5)(u-1.5)(u-4)$, $V=0.66$, $b=1$, $a=1$, $c=2$, $d=0$, $D_1=1,D_2=0.1$, 
    and $p=0$. 
    $500 \leq t \leq 1200$.
    Initial conditions of both $u_i(0)$ and $\frac{d u_i}{d t}(0)$ are given with a uniform random number between $-2$ and $2$.
    }  
    \label{fig:noiseless}
\end{figure}

Regarding the effect of noise, the noise term is necessary for the reproduction of the fractal spatiotemporal patterns in the previous model\cite{yamazaki2006pattern}. 
However, the present model does not need any noise term 
in the time evolution of $u$ to reproduce the patterns as shown in Fig.\ref{fig:modelPattern}, 
although we added the noise in our numerical calculation for stabilizing the state B.
Actually, Fig.\ref{fig:noiseless} shows an example of the spatiotemporal pattern 
obtained from eq.(\ref{eq:model4}) without noise term; 
we considered the randomness in $u$ only for the initial condition.
This result suggests that the fractal spatiotemporal patterns 
originate not from stochasticity but from chaoticity.

Finally, we note that the fractal patterns as shown in Figs.\ref{fig:modelPattern}(c) and \ref{fig:noiseless}
are reproducible by Bonhoffer-van der Pol type equation, and other reaction-diffusion equations~\cite{hayase1997collision,hayase1998sierpinski,chate1999forcing,hayase2000replicating}. 
The relationship between our present model and these models can be found by transforming our model into the 2-component activator-inhibitor system 
as the Li\'{e}nard system by introducing a new parameter
$w_i\equiv\frac{du_i}{dt}+bu_i-a'\arctan (c'(u_i-d))-D_2(u_{i+1}+u_{i-1}-2u_{i})$:
\begin{eqnarray}\label{lienard1}
    \frac{d u_i}{d t}&=& -w_i-bu_i+a'\arctan (c'(u_i-d))+D_2(u_{i+1}+u_{i-1}-2u_{i})\\
    \label{lienard2}
    \frac{d w_i}{d t}&=& \left.\frac{d U}{d u}\right|_{u=u_i}+bV-D_1(u_{i+1}+u_{i-1}-2u_{i}),
\end{eqnarray}
where $a'=a/\sqrt{c}, c'=\sqrt{c}$.
If we change the term $\frac{d U}{d u}+bV$ in eq.(\ref{lienard2}) to $u$, 
then the reaction terms of eqs.(\ref{lienard1}) and (\ref{lienard2}) 
correspond to those of the Bonhoffer-van der Pol type equation\cite{hayase1997collision}.

The presence of such correspondence also suggests that the fractal patterns emerge chaotically.

\textit{Conclusion}: 
In this letter, we have proposed a new model for pattern formation in peeling adhesive tapes
which focuses on the equation of motion for the displacement of adhesives in the peel front.
We introduce spatial interaction terms without the asymmetric interaction, thus, it can be interpreted as a mechanical model of the peeling front. Also, we introduced the new interaction terms which are introduced from the rheological property of the adhesives, viscoelasticity.

The model reproduces the dynamical and statistical properties of the spatiotemporal patterns which consist of the two different peel states.
The previous experimental results also support our numerical results. 
Those statistical properties can also be reproduced from the previous models, because our model and the previous models are related mathematically. We will report the discussion of such mathematical properties of the models in near future.
Note that the spatiotemporal pattern obtained from the present model is seemingly relevant to directed-percolation universality class~\cite{hinrichsen2000dp}.
Detailed discussions of this relevancy will also be reported in near future.
\ \\
\section*{Acknowledgments}
We appreciate Prof. H. Nakao for valuable discussions. 
We acknowledge JST CREST JP-MJCR1913 for financial support.

\ \\
\ \\

\end{document}